# Transformation of envelope solitons on a bottom step

G. Ducrozet[1], A.V. Slunyaev[2,3*)], Y.A. Stepanyants[4,5]

[1]Ecole Centrale Nantes, LHEEA Laboratory (ECN and CNRS), 1 rue de la Noë, 44300 Nantes, France;
[2]National Research University – Higher School of Economics, 25 Bolshaya Pechorskaya Street, Nizhny Novgorod, 603950, Russia;
[3]Institute of Applied Physics of the Russian Academy of Sciences, 46 Ulyanov Street, Box-120, Nizhny Novgorod, 603950, Russia;
[4] Department of Applied Mathematics, Nizhny Novgorod State Technical University, n.a. R.E. Alekseev, 24 Minin Street, Nizhny Novgorod, 603950, Russia;
[5]School of Sciences, University of Southern Queensland, Toowoomba, West St., QLD, 4350, Australia.

**Abstract**

In this paper we study the transformation of surface envelope solitons travelling over a bottom step in water of a finite depth. Using the transformation coefficients earlier derived in the linear approximation, we find the parameters of transmitted pulses and subsequent evolution of the pulses in the course of propagation. Relying on the weakly nonlinear theory, the analytic formulae are derived which describe the maximum attainable wave amplitude in the neighbourhood of the step and in the far zone. Solitary waves may be greatly amplified (within the weakly nonlinear theory formally even without a limit) when propagating from relatively shallow water to the deeper domain due to the constructive interference between the newly emerging envelope solitons and the residual quasi-linear waves. The theoretical results are in a good agreement with the data of direct numerical modelling of soliton transformation. In particular, more than double wave amplification is demonstrated in the performed simulations.

*)Corresponding author: Slunyaev@appl.scinnov.ru

# 1. Introduction

The problem of wave transformation in the water of a variable bathymetry has a long history and still remains one of the topical problems in the classical hydrodynamics with applications to physical oceanography, hydraulic theory, and other fields of practical hydrodynamics. In the case of the simplest bathymetry with a single bottom step shown in Fig. 1, the problem was intensively studied in the linear approximation in 1950– 1960-ies; the review of the results obtained can be found in Refs. (Kurkin et al., 2015a; 2015b). The important role in the wave transformation plays evanescent modes a countable number of which arise in the neighborhood of a bottom step.

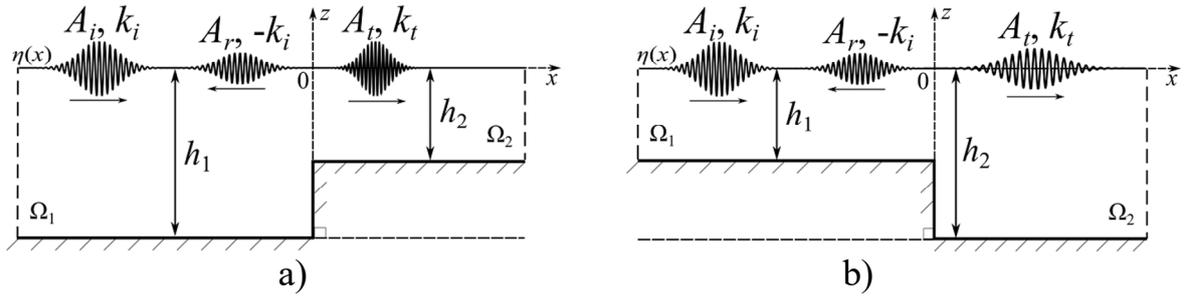

Fig. 1. Sketch of the problem with two configurations of incident and transformed wave trains.

The efficient way to determine the transformation coefficients (the transmission $T$ and reflection $R$ coefficients) for the quasi-sinusoidal surface wave have been found after a thorough problem study by many researchers (see Refs. (Kurkin et al., 2015a; 2015b) and references therein). Besides of the determination of the transformation coefficients, the coefficients of excitation of evanescent modes have been obtained in the cited papers too; all these require solution of an infinite set of algebraic equations, in general. However, the set of equations can be truncated at a finite number of modes for the practical application; this allows one to determine the excitation coefficients for the first $N_m$ evanescent modes. The accuracy of such an approach can be very high if the appropriate number of modes are taken into consideration because the coefficients quickly decrease with the mode number. The choice of $N_m$ depends on the required accuracy of the wave field representation in the neighborhood of a bottom step; in particular, the smoothness of the surface elevation depends on $N_m$. In Ref. (Kurkin et al., 2015a) the excitation coefficients were calculated numerically up to $N_m = 500$ evanescent modes.

In the meantime, the values of the transformation coefficients are not so sensitive to the choice of the number of evanescent modes. In particular, as was shown by Miles (1967), the transformation coefficients of surface waves can be evaluated with the 95% accuracy by neglecting all the evanescent modes. A simple heuristic approach to the evaluation of the transformation coefficients was suggested in Refs. (Giniyatullin et al., 2014; Kurkin et al., 2015a) that provides the same or even better accuracy as the Miles' method but contains much simpler formulae. The expressions for the transformation coefficients naturally reduce to the classical Lamb formulae (Lamb, 1933) in the long-wave approximation:

$$T = \frac{2}{1+c_2/c_1} = \frac{2}{1+\sqrt{h_2/h_1}}, \quad R = \frac{1-c_2/c_1}{1+c_2/c_1} = \frac{1-\sqrt{h_2/h_1}}{1+\sqrt{h_2/h_1}}, \tag{1}$$

where $c_{1,2} = \sqrt{gh_{1,2}}$ are the speeds of long linear waves, $g$ is the acceleration due to gravity, and $h_{1,2}$ are the water depth on each side of the bottom step (see Fig. 1).

The nonlinear effects in the shoaling zone play a profound role leading to the strong changes of the water wave spectra and individual wave shapes. The problem of the nonlinear wave transition above fast depth variations has received much interest in the recent years, particularly due to the realization of the anomalous wave statistics in the vicinity of the abrupt depth change. In particular, the probability of appearance of waves of extremely high amplitudes noticeably increases over a shoal zone (Sergeeva et al., 2011; Trulsen et al., 2012; Zeng & Trulsen, 2012); this is a matter of obvious practical importance in the context of risk assessment posed by anomalous waves in coastal zones. The understanding of this problem turns out to be complicated because many physical effects interplay whereas the traditional simplifying assumptions are not always efficient. The comprehensive overviews of the performed research were given in the recent publications by Trulsen et al. (2020) and Li et al. (2021a, b) (see the summarizing diagrams in Figs. 1 in the cited papers. Other results of recent numerical simulations may be found in (Zheng et al., 2020; Zhang & Benoit, 2021). The most fascinating effect is that the surface elevation can have a local maximum of skewness and kurtosis above the shallower part of the shoal. The bar-profile or step-profile bottoms were considered in the cited works, where the waves travelled from the deeper to shallower water. The shallower domains usually corresponded to the essentially small-depth conditions, $kh < 1$, where $k$ is the wavenumber.

Though the trend in the research seems to move towards the experimental and fully nonlinear simulations of irregular waves, nevertheless, a qualitative understanding of the working physical mechanisms as well as simplified but efficient models, capable to provide quantitative estimations, are strongly required. The second-order theory for wave groups travelling over a bottom step was developed in (Li et al., 2021a, b) and then, validated numerically and experimentally. The consideration of the model groups allowed the authors to obtain a detailed theoretical description of nonlinear processes which underlie the strong deformation of the wave envelope and generation of new wave groups. Similar to other works, in (Li et al., 2021a,b) the focus was made on such cases when waves experienced a fast transition to the shallower domains.

In the present work, we investigate the transformation of weakly-nonlinear wave trains in the form of envelope solitons that pass over the stepwise depth change. We show that an emerged solitary wave group can be amplified a few times when it travels from a sufficiently shallow domain to a much deeper domain. In Section 2, we use a combination of linear and weakly nonlinear theories for slowly modulated waves to obtain the analytic description of the main wave characteristics, including the estimation of the wave amplitude after passing the bottom step. The soliton transmission coefficient is thoroughly analyzed in Section 3, where the regimes of the maximum wave enhancement are determined. The effect of wave group amplification due to the generation of new envelope solitons and interaction with them is studied in Section 4 by means of the direct numerical simulation within the weakly nonlinear model. The analytic description of the maximum attainable wave amplitude is proposed. The discovered effects of the envelope soliton transformation are further verified in Section 5 by means of direct numerical simulation of the primitive set of the hydrodynamic equations for the potential motion (will be called hereafter the Euler equations) by means of the High Order Spectral Method (HOSM). The accuracy of the analytic estimations is examined in this section as well. We are completing the paper with concluding remarks in Section 6.

**2. The analytic theory for the transformation of an envelope soliton at the bottom step**

We consider weakly-nonlinear quasi-monochromatic wave groups propagating in the domains $\Omega_1$ and $\Omega_2$ as shown in Fig. 1 far from the bottom step. Such wave groups can be described by the nonlinear Schrödinger (NLS) equation (see, e.g., (Djordjevic & Redekopp, 1978; Mei et al., 2005)):

$$i\left(\frac{\partial A_{1,2}}{\partial x} + \frac{1}{C_{1,2}}\frac{\partial A_{1,2}}{\partial t}\right) + \beta_{1,2}\frac{\partial^2 A_{1,2}}{\partial t^2} + \alpha_{1,2}\left|A_{1,2}\right|^2 A_{1,2} = 0, \qquad (2)$$

where $A_j(x,t)$ are the complex amplitudes of the incident and transmitted wavetrains in the domains $\Omega_1$ ($j = 1$) and $\Omega_2$ ($j = 2$), respectively. The fate of the reflected wave will not be of our interest, but the reflected wave will be taken into account in the wavetrain transformation at the step. According to the NLS theory, the surface displacements $\eta_j(x, t)$ are specified through the relations:

$$\eta_j(x,t) = \text{Re}\left[A_j \exp(i\omega t - ik_j x)\right], \quad j = 1, 2. \tag{3}$$

The carrier frequency $\omega$ and wavenumbers $k_j$ are related through the dispersion relations for the finite-depth water in each domain:

$$\omega = \sqrt{gk_j \sigma_j}, \quad \sigma_j = \tanh(k_j h_j), \quad j = 1, 2, \tag{4}$$

where $h_j$ are the water depths in the corresponding domains. Note that in the stationary case when the medium parameters do not vary in time, the frequency conserves; therefore, all waves participating in the transformation (incident, transmitted and reflected) possess the same frequency. The wave group velocities, $C_j$, in Eq. (2), as well as the dispersion and nonlinear coefficients, $\beta_j$, and $\alpha_j$, respectively, depend on the local water depth and are given by the following expressions (see, e.g., Zeng & Trulsen, 2012):

$$C = \frac{\partial \omega}{\partial k} = \frac{g}{2\omega}\left[\sigma + kh(1-\sigma^2)\right] \tag{5}$$

$$\beta = \frac{1}{2C^3}\frac{\partial^2 \omega}{\partial k^2} = \frac{1}{2C\omega}\left[1 - \frac{C_{LW}^2}{C^2}(1-kh\sigma)(1-\sigma^2)\right], \tag{6}$$

$$\alpha = \frac{\omega k^2}{16\sigma^4 C}\left(9 - 10\sigma^2 + 9\sigma^4 - \frac{2\sigma^2 C^2}{C_{LW}^2 - C^2}\left(4\frac{C_p^2}{C^2} + 4\frac{C_p}{C}(1-\sigma^2) + \frac{C_{LW}^2}{C^2}(1-\sigma^2)^2\right)\right), \tag{7}$$

where $C_p = \frac{\omega}{k}$, is the phase velocity, and $C_{LW} = \sqrt{gh}$ is the velocity of long linear waves. The subscripts denoting the number of the domain are omitted here for the sake of brevity. It is well known that the coefficients $C$ and $\beta$ are positive at any depth, whereas the coefficient $\alpha$ is negative for $kh < 1.363$ and is positive otherwise. Further it is convenient to use the following dimensionless wavenumbers in the respective water domains: $\kappa_1 = k_1 h_1$, and $\kappa_2 = k_2 h_2$.

## 2.1. Transformation of a wavenumber

Due to the frequency conservation in the process of wave transformation, one can consider the relation $\omega^2/g = k\tanh(kh) = \text{const}$ for the function $k = k(h) > 0$. Then, it can be straightforwardly shown that $dk/dh < 0$ and $d(kh)/dh > 0$. Hence, the wavenumber of the transmitted wave increases when the incident wave arrives from a deeper domain; in the opposite case, when the incident wave comes from a shallower domain, the wavenumber of the transmitted wave decreases. Due to the frequency conservation in a stationary medium, we can find the relationship between the wavenumbers of carrier waves in the domains $\Omega_1$ and $\Omega_2$ through the equation:

$$k_2 \tanh(kh_2) = k_1 \tanh(kh_1). \tag{8}$$

The ratio of the wavenumbers as a function of the depth jump is shown in Fig. 2. In the limiting case of the long-wave approximation, when $\kappa_1 \to 0$ and $\kappa_2 \to 0$, Eq. (8) simplifies and reduces to the well-known Lamb's formula $k_2/k_1 = (h_1/h_2)^{1/2}$ (Lamb, 1933). A similar plot was presented in the paper (Giniyatullin et al., 2014) where there was a misprint in the definition of $q = k_2 h_2$, that should be $q = k_2 h_1$.

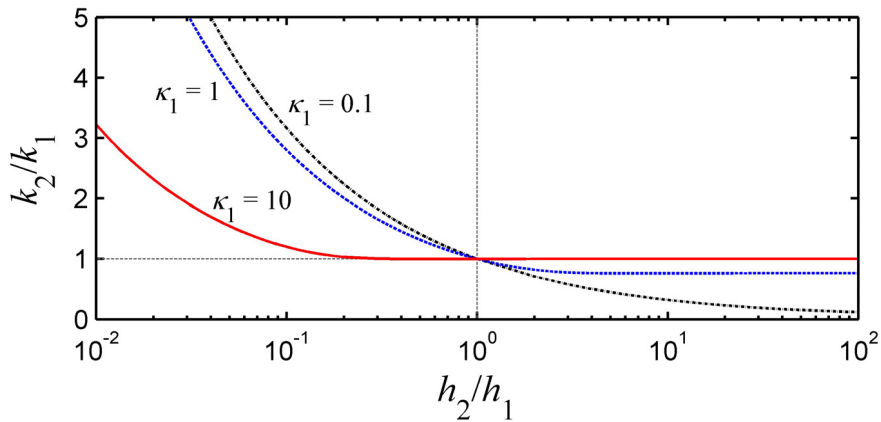

Fig. 2 (color online). The dependence of the wavenumber ratios as the function of the depth jump $h_2/h_1$ for the different values of the normalized wavenumber of incident wave, $\kappa_1 = k_1 h_1$. Hereafter the numbers at the lines correspond to the order of curves in the legend.

As one can see from Fig. 2, the change of the wavenumber can be very big when a wave enters the shallower region, i.e., when $h_2/h_1 \to 0$. In another limit, all curves asymptotically approach some constant values $(k_2/k_1)_{lim}$ which depends on $\kappa_1$:

$$(k_2/k_1)_{lim} = \lim_{h_2/h_1 \to \infty} \frac{\tanh k_1 h_1}{\tanh k_2 h_2} = \tanh \kappa_1; \tag{9}$$

this is illustrated by Fig. 3. If $\kappa_1 \gg 1$, then the hyperbolic tangents in Eq. (9) turn to unity, and the wavenumber of the transmitted wave becomes equal to the wavenumber of the incident wave. Therefore, the dependence $(k_2/k_1)_{lim}$ asymptotically approaches unity when $\kappa_1 \to \infty$ (see, for example, the case $\kappa_1 = 10$ in Fig. 2). Practically $k_2$ becomes equal to $k_1$ when $\kappa_1 > 3$, frequently considered as the deep-water threshold for water waves.

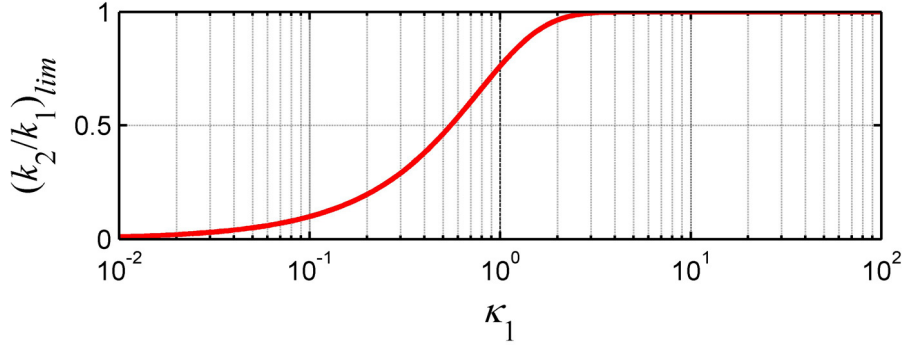

Fig. 3 (color online). The dependence of $(k_2/k_1)_{lim}$ on the dimensionless wavenumber $\kappa_1$ of the incident wave.

### 2.2. Transformation of an envelope soliton

When a wave train passes over the bottom step, it splits into the transmitted and reflected pulses; besides, the evanescent modes are generated in the neighborhood of the step. The amplitude of the transferred wave can be determined through the transformation coefficient $T(\kappa_1, \kappa_2)$, which depends on the depth drop (see, e.g., (Kurkin et al., 2015a; 2015b) and references therein). Adopting this description to the case of narrow-band spectrum waves, we relate the solution in the domain $\Omega_2$ with the transmitted wave by the boundary condition at $x = 0$ (the position where the bottom jumps from $h_1$ to $h_2$):

$$A_2(x=0,t) = TA_1(x=0,t). \qquad (10)$$

Let us define the scaled amplitude and coordinate of the transmitted wavetrain as:

$$\tilde{A}_2(\tilde{x},t) = SA_2(x,t), \qquad S = \sqrt{\frac{\alpha_2 \beta_1}{\alpha_1 \beta_2}}, \qquad \tilde{x} = x\frac{\beta_2}{\beta_1}. \qquad (11)$$

Then, the dynamics of the transmitted wave train is described by the following NLS equation for $\tilde{x} \geq 0$:

$$i\left(\frac{\partial \tilde{A}_2}{\partial \tilde{x}} + \frac{1}{\tilde{C}_2}\frac{\partial \tilde{A}_2}{\partial t}\right) + \beta_1 \frac{\partial^2 \tilde{A}_2}{\partial t^2} + \alpha_1 |\tilde{A}_2|^2 \tilde{A}_2 = 0, \qquad \tilde{C}_2 = C_2\frac{\beta_2}{\beta_1}. \qquad (12)$$

Equation (12) for $\tilde{A}_2(\tilde{x},t)$ differs from Eq. (2) with $j = 1$ in the advective part only; the envelopes $A_1$ and $\tilde{A}_2$ evolve along the x-axis similarly but with the different speeds. Note that the relations (11) imply that the parameter $\alpha$ has the same sign on both sides of the bottom step (we recall that according to Eq. (6), the coefficient $\beta$ is positive). The matching condition at $x = 0$ specifies the Cauchy problem for $\tilde{A}_2(\tilde{x},t)$ to be solved for Eq. (12); according to Eqs. (10) and (11) the condition reads:

$$\tilde{A}_2(\tilde{x}=0,t) = \mu A_1(x=0,t), \quad \mu \equiv TS. \tag{13}$$

Let us consider in detail the case when $\alpha_1\beta_1 > 0$ and $\alpha_2\beta_2 > 0$, hence the modulation (Benjamin–Feir) instability (see, e.g., in (Mei et al., 2005)) can develop in both domains $\Omega_1$ and $\Omega_2$; this secure the possibility of existence of envelope solitons. We assume that the incident wave represents an envelope soliton of the NLS equation with the real amplitude $a$ and the carrier frequency $\omega$:

$$A_1(x,t) = a \exp\left(\frac{i}{2}a^2\alpha_1 x\right) \text{sech}\left[a\sqrt{\frac{\alpha_1}{2\beta_1}}\left(t - \frac{x}{C_1}\right)\right]. \tag{14}$$

Then, the boundary condition at $\tilde{x} = 0$ for Eq. (12) has the form of the sech-shape envelope with the amplitude factor $\mu$:

$$\tilde{A}_2(\tilde{x}=0,t) = \mu a \, \text{sech}\left(\sqrt{\frac{\alpha_1}{2\beta_1}}\, a t\right). \tag{15}$$

In the course of propagation in the domain $\Omega_2$, the transmitted pulse will experience a nonlinear evolution. The problem of pulse evolution, when the initial condition has the sech-shape, was studied in (Slunyaev et al., 2017) both by means of the exact analytic solution obtained by Satsuma & Yajima (1974) and numerically. According to this solution, the number of secondary envelope solitons $N$ which emerge from the sech-type initial condition, and their amplitudes, $\tilde{a}_n$ are given by the following expressions:

$$N = E\left(\mu + \frac{1}{2}\right), \quad \tilde{a}_n = 2a\left(\mu - n + \frac{1}{2}\right), \quad n = 1, 2, \ldots, N \tag{16}$$

where $E(\cdot)$ stands for the integer part of the argument. When a soliton group passes over a step, it keeps its shape but changes the amplitude. As follows from Eq. (16), if the parameter $\mu$ ranges from 0.5 to 1, then at most one soliton can emerge from the transmitted pulse. But if $\mu < 0.5$, the transmitted pulse completely disintegrates into a quasi-linear dispersive wavetrain. When $\mu > 1$, the soliton amplitude increases, and other secondary envelope solitons can emerge from the transmitted pulse if $\mu > 1.5$. In particular, two solitons asymptotically emerge when $1.5 < \mu < 2.5$. The unlimited number of solitons can formally emerge if $\mu \to \infty$; the amplitudes of generated solitons increase infinitely too. The validity of the approximate solution given by Eq. (16) for $\tilde{a}_n$ was verified in the numerical simulations of the Euler equations in the case of infinitely deep water (Slunyaev et al., 2017). In that paper, the range of parameters $0.2 \leq \mu \leq 1.2$ was examined for the emergence of envelope solitons up to the relatively high steepness $k_1 a \approx 0.25$.

Solution (16) is written for the auxiliary function $\tilde{A}_2(\tilde{x}, t)$ and should be rewritten for the original problem scales with the help of Eq. (11). Then, the soliton amplitudes $a_n$ in the second domain $\Omega_2$ are determined by the following equations:

$$a_n = 2\frac{a}{S}\left(\mu - n + \frac{1}{2}\right), \quad n = 1, 2, \ldots N, \quad N = E\left(\mu + \frac{1}{2}\right), \quad \mu = ST. \tag{17}$$

Note that solutions (16) and (17) provide the same number of envelope solitons, but their amplitudes can be remarkably different. In particular, for the fixed coefficient $T$ in the limit $S \to \infty$, the number of solitons $N$ also goes to infinite, but their amplitudes in the domain $\Omega_2$ are limited from above by the value $a_n < 2Ta$. This condition realizes, in particular, when the incident wavetrain travels from the region of the minimal permissible depth where $\kappa_1 \approx 1.363$ to the deeper region (the nonlinear coefficient $\alpha_1$ vanishes in the region with the critical depth $\kappa_1 = 1.363$). In general, the amplitude of the leading envelope soliton $a_1$ is greater or smaller than the amplitude of the incident envelope soliton $a$ depending on the relation between the parameters $S$ and $T$; $a_1 > a$, if

$$S > \frac{1}{2T - 1} \tag{18}$$

and $a_1 < a$, in the opposite case. Consequently, for the given $S$ the soliton amplitudes increase if $T > 1$ and decrease if $T < 1$.

If more than one soliton emerges in the domain $\Omega_2$, then all of them possess the same carrier frequency and are located in the same point which drifts with the group velocity $C_2$. Therefore, within the NLS theory they remain coupled in the course of evolution and exhibit breathing-type envelopes. Besides the new solitons, a dispersive wave train can be generated which spreads with time.

Therefore, within the employed assumptions which simplify the problem, the parameter $\mu$ plays a crucial role; it determines both the number of envelope solitons which arise in the transmitted wave field and their amplitudes. We refer further to the parameter $\mu$ as to the main parameter determining the soliton transmission. Below we consider the dependence of $\mu$ on the problem parameters in detail.

## 3. Coefficients of soliton transmission

As was abovementioned, we consider such a case when 'bright' envelope solitons can exist in both domains $\Omega_1$ and $\Omega_2$, i.e. when $\alpha_j\beta_j > 0$, $j = 1, 2$. In this case, a plane wave is affected by the modulation instability (see, e.g., (Zakharov & Ostrovsky, 2009; Mei et al, 2005)). These inequalities restrict water depths from below, $\kappa_j > 1.363$. Functions $\sigma_j(\kappa_j)$ in Eq. (4) are limited from below, $\sigma_j > 0.877$, under these conditions; they approach a unity exponentially quickly when $\kappa_j$ increases. As the result, the carrier wavenumber does not change too much; as follows from Eq. (8), $(k_2 - k_1)/k_1 < \coth(1.363) - 1 \approx 0.14$. The smaller the water depth change, the less is the variation of the carrier wavenumber.

According to results obtained in (Giniyatullin et al., 2014; Kurkin et al., 2015), the transmission coefficient $T$ can be calculated with a rather high accuracy using the empirical formula:

$$T = \frac{2C_1}{C_1 + C_2}, \qquad (19)$$

where $C_1$ and $C_2$ are the group velocities of the incident and transmitted waves, respectively. This formula was validated by comparison with the results of direct numerical simulation within the set of primitive hydrodynamic equations. In what follows, we use this formula to calculate the transmission coefficient $T$ for the envelope solitons assuming that they have a narrow spectrum. We also assume that the characteristic length of the process of the wave transmission over a bottom step is much shorter than the characteristic distance of manifestation of nonlinear effects, i.e. the process of transformation can be described within the linear theory. Then, the coefficient $T$ depends on the parameters $\kappa_1$ and $\kappa_2$ only.

The isolines of the coefficient $T$ in Eq. (19) as a function of $\kappa_1$ and the depth ratio $h_2/h_1$ are plotted in Fig. 4. It can be readily shown from Eq. (5), that the group velocity as a function of depth $h$ for a fixed $\omega$ is always positive and attains maximum at the point $\kappa_1 \approx 1.200$ which is the root of the equation $\kappa_1 = \coth \kappa_1$. Therefore, in the situation of our interest, $\kappa_1 > 1.363$, the group velocity decreases when $h$ increases, and $T$ grows in this case. Consequently, waves increase in amplitude ($T > 1$) when they travel to the deeper domain and decrease ($T < 1$) in the opposite case. Note that this conclusion is valid to sufficiently deep water, $kh > 1.2$. In shallower basins, $kh < 1.2$, the change of the wave amplitude over a bottom step will be opposite as discussed in [Kurkin et al, 2015a]. The local minimum of function $T$ for a given $\kappa_1$ should be reached at $\kappa_2 \approx 1.2$, which corresponds to the defocusing regime in the domain $\Omega_2$, when $\alpha_2\beta_2 < 0$ (the white area in Fig. 4). As one can see from Fig. 4, the values of $T$ noticeably differ from the unity in the rather localized ranges of parameters: they are significantly greater than one in the range of small depths $\kappa_1 < 3$ when $h_2 > h_1$; and are significantly smaller when the depth in the second domain tends to the limit $\kappa_2 = 1.363$ (the lower edge of the colored area in Fig. 4). In general, for $\kappa_1 > 1.363$ and $\kappa_2 > 1.363$ the range of coefficients $T(\kappa_1, \kappa_2)$ is confined within the interval from $T(\infty, 1.363) \approx 0.88$ to $T(1.363, \infty) \approx 1.12$. The linear wave amplification could be much larger if the basins were shallower [Kurkin et al, 2015a].

As the coefficient $T$ does not vary much, the isolines of the functions $S$ and $\mu = TS$ look almost alike. One can readily confirm that the ratio $\alpha/\beta$ grows when the depth increases, hence $S > 1$ if $h_2 > h_1$ and $S < 1$ in the opposite case $h_2 < h_1$, similar to the transmission coefficient $T$. The contour plot in Fig. 5 illustrates the dependence of the parameter $\mu$ on $\kappa_1$ and $h_2/h_1$. The panel (a) pertains to the case when a wavetrain penetrates from the shallower to deeper domain, and the panel (b) pertains to the opposite case.

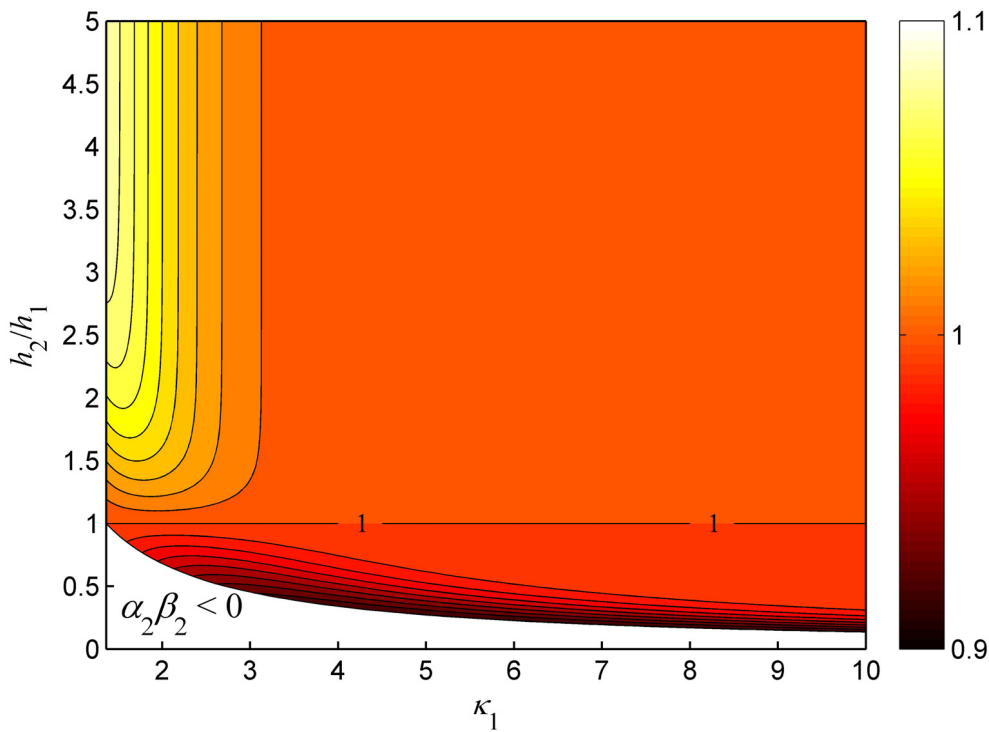

Fig. 4 (color online). Isolines of the linear transition coefficient $T$ as per Eq. (19). The uncolored area corresponds to the defocusing domain $\Omega_2$ where $\kappa_2 < 1.363$.

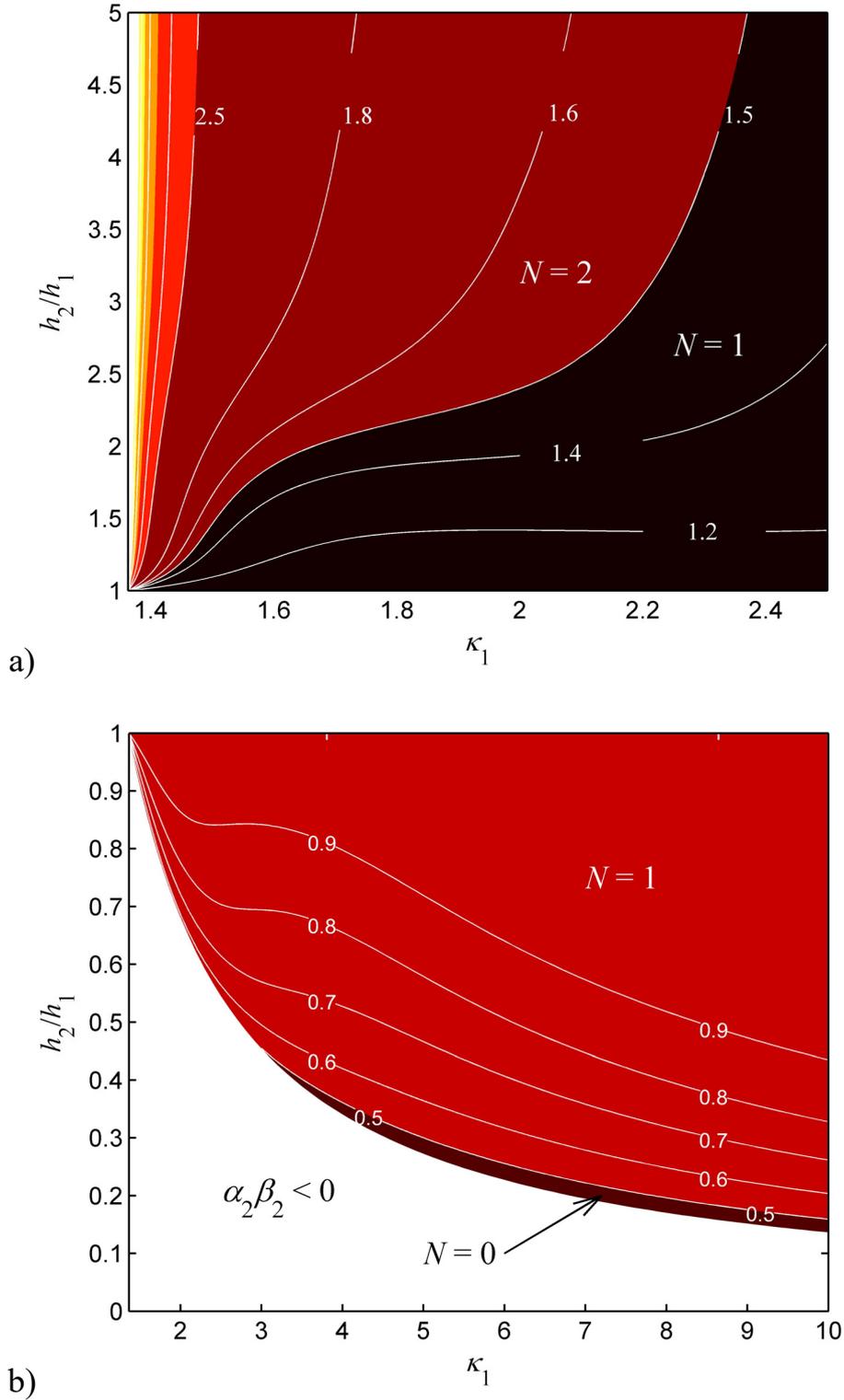

Fig. 5 (color online). Contour plots of the transmission parameter $\mu = TS$ when an envelope soliton propagates to deeper (a) and shallower (b) domains (note the different limits of the axes). Different colors show the domains where different number $N$ of solitons emerge in the transmitted wave filed. The white area in the panel (b) corresponds to the condition when the waves in the shallower domain are modulationally stable, $\kappa_2 < 1.363$.

If an envelope soliton travels from the shallow to deep water, then $\mu > 1$; the amplitude of the transmitted soliton increases, and the number of secondary solitons can be greater than 1, see Fig. 5a. A bigger depth drop is required to cause more significant modification in the wavetrain. Soliton parameters are the most sensitive to the abrupt deepening, when an incident soliton arrives from the relatively shallow water where $\kappa_1 < 2$. Different colors in Fig. 5a show the domains where the different number $N$ of secondary envelope solitons can emerge in the transmitted wavefield as per Eq. (17).

As the ratio $\alpha/\beta$ grows when the depth increases, in the limiting case $kh \to \infty$, the maximum value of this ratio is attained:

$$\sqrt{\frac{\alpha}{\beta}} \xrightarrow{kh \to \infty} \sqrt{\frac{\alpha_\infty}{\beta_\infty}} = \frac{\omega^3}{g}. \tag{20}$$

Then, the maximum value of $\kappa_1$, when a second envelope soliton can emerge from an incoming soliton, corresponds to the case when the depth changes from $h_1$ to infinite. It is prescribed by the condition, which yields the threshold value of $\kappa_1$:

$$\mu(\kappa_1, \kappa_2 = \infty) > 1.5 \quad \text{when} \quad \kappa_1 < \kappa_1^{(2\,\text{sol})} \approx 2.859. \tag{21}$$

When the soliton propagates initially in deeper water, $\kappa_1 > \kappa_1^{(2\,\text{sol})}$, new envelope solitons cannot emerge. Three solitons emerge when $\mu(\kappa_1, \kappa_2=\infty) > 2.5$; this requires $\kappa_1 < \kappa_1^{(3\,\text{sol})} \approx 1.820$. If the depth in the second domain of water is finite, the values of $\kappa_1^{(2\,\text{sol})}$ and $\kappa_1^{(3\,\text{sol})}$ further decrease.

Similarly, the condition

$$\mu(\kappa_1 = \infty, \kappa_2) < 0.5 \quad \text{when} \quad \kappa_2 < \kappa_2^{(\text{destr})} \approx 2.149 \tag{22}$$

provides the upper limit on the depth behind the bottom step, when the transmitted envelope soliton gets completely destroyed. With the use of Eq. (8), the critical condition (22) on the depth in the second zone can be reformulated in terms of the wavenumber of the incident envelope soliton $k_1 h_2 \approx 2.091$. If the depth behind the step is deeper, then an incident envelope soliton will recover as the transmitted soliton after the depth change but with the reduced amplitude. If the depth prior to the step is finite, the transmitted pulse disintegrates at the shallower domain $\Omega_2$.

The fact that the transmission coefficient $\mu$ is monotonic with respect to the depth change $h_2/h_1$ for the given $\kappa_1$ leads to the conclusion that the soliton amplitude (the leading soliton amplitude, if several are generated) always grows when $h_2 > h_1$ and diminishes when $h_2 < h_1$. According to Eq. (17), the maximum amplification of the envelope soliton amplitude, $a_1/a$, is restricted by the value of $2T$ in the limit of large $\mu$, when the number of generated secondary solitons is big. Within the NLS theory, the number $N$ of generated solitons and the amplification factors $a_n/a$ do not depend on the amplitude of the incident pulse. The solution for the soliton amplitudes according to Eqs. (17) and (19) is shown for a few values of $\kappa_1$ in Fig. 6. It turns out that the amplitude of the leading soliton $a_1$ can exceed the limit of double amplitude of the incident soliton, $2a$, only when the soliton comes to a very deep domain from extremely shallow water, $\kappa_1 < 1.39$ (not shown in the figure). In the case when the soliton travels from deep to shallower water, either at most one soliton of reduced amplitude emerges in the transmitted wavefield or the transmitted pulse completely disintegrates.

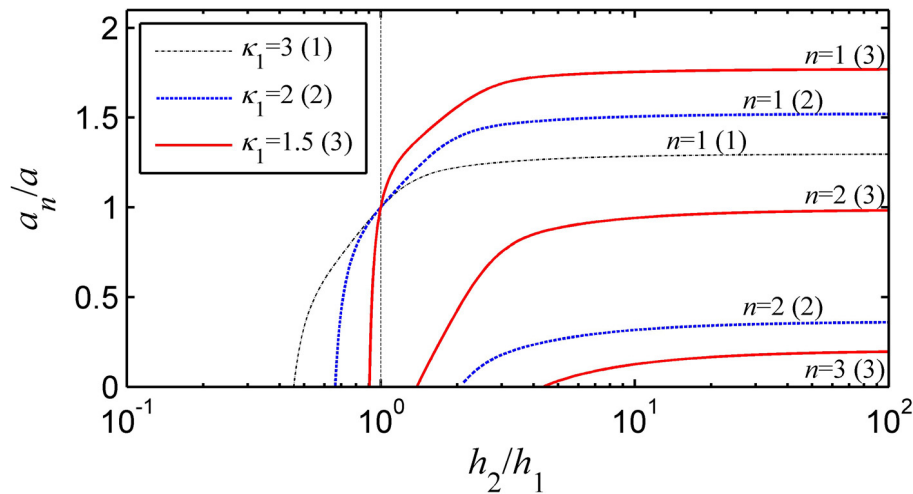

Fig. 6 (color online). Amplitudes $a_n$ of envelope solitons generated in the domain $\Omega_2$, normalized by the amplitude of the incident soliton $a$, according to Eqs. (17) and (19) for different values of $\kappa_1$. The numbers in brackets indicate the cases.

For the waves traveling from deep to shallower water, the parameter $\mu < 1$, as shown in Fig. 5b, and therefore, only one envelope soliton can emerge in the domain $\Omega_2$ (see the area for $N = 1$). The amplitude of the transmitted soliton is smaller than the amplitude of the original incident soliton arriving from the domain $\Omega_1$. The transmitted pulse can completely disintegrate if $\mu < 0.5$, see the domain marked with $N = 0$ below the isoline $\mu = 0.5$. The white area in Fig. 5b shows the range of parameters where envelope solitons cannot exist in the shallower domain $\Omega_2$ because $\kappa_2 < 1.363$. Note the nonmonotonic shape of isolines in the intermediate depth interval where $2 < \kappa_1 < 4$.

# 4. Numerical simulation of soliton transformation within the nonlinear Schrödinger equation

In Sec. 3, the transformation of an envelope soliton on a bottom step was analyzed in terms of soliton amplitudes in the transmitted domain $\Omega_2$. This provides us with the asymptotic solution far from the bottom step when $x \to +\infty$. Let us consider the transient dynamics which occurs in the domain $\Omega_2$ right after the step in more detail. To this end, the NLS equation in this domain, Eq. (2) with $j = 2$, was solved numerically. The parameters of the incident envelope soliton were chosen such that the steepness was $k_1 a = 0.12$. In what follows, the results will be presented in the following dimensionless spatial and temporal coordinates:

$$\xi_2 = \frac{k_2}{2\pi} x, \qquad \tau_2 = \frac{\omega_2}{2\pi}\left(t - \frac{x}{C_2}\right). \tag{23}$$

These variables represent the number of wavelengths and wave periods respectively, in the domain $\Omega_2$. A different choice of the steepness will cause different characteristic scales of the evolution but will not change the dynamics qualitatively.

Let us consider first the case which is usually called the 'intermediate' depth, when $\kappa_1 = 2$. The situations when solitons recover in the transmitted domain $\Omega_2$ are illustrated in Fig. 7 where the transitions from the deeper to shallower domains occurs ($h_2/h_1 = 0.8$, left frames (a) and (c)) and vice versa – from the shallower to the deeper domains ($h_2/h_1 = 1.5$, right frames (b) and (d)). The spatio-temporal evolution of envelopes of the transmitted wavetrains $|A_2|$ are shown in Figs. 7a and 7b; while Figs. 7c and Fig. 7d show by solid lines the evolution of $\max|A_2|$ as function of distance. The amplitudes of envelope pulses experience quasi-periodic oscillations in both cases. The horizontal blue dashed lines in Figs. 7c and 7d show the scaled amplitudes $a_1/a$ of envelope solitons in the transmitted domain $\Omega_2$ calculated according to Eqs. (17) and (19). It is clearly seen that $\max|A_2|$ gradually converges to the theoretical value when $\xi_2 \to \infty$. When the water depth decreases, the maximum of the envelope pulse drops down at the step and then experiences decaying oscillations gradually approaching the amplitude of the envelope soliton $a_1 < a$ (Fig. 7c). In the opposite case when the water depth increases, the maximum of the envelope pulse increases right after the depth change due to $T > 1$; then it further grows and finally experiences decaying oscillations around the theoretically predicted amplitude of the envelope soliton, $a_1 > a$ (Fig. 7d).

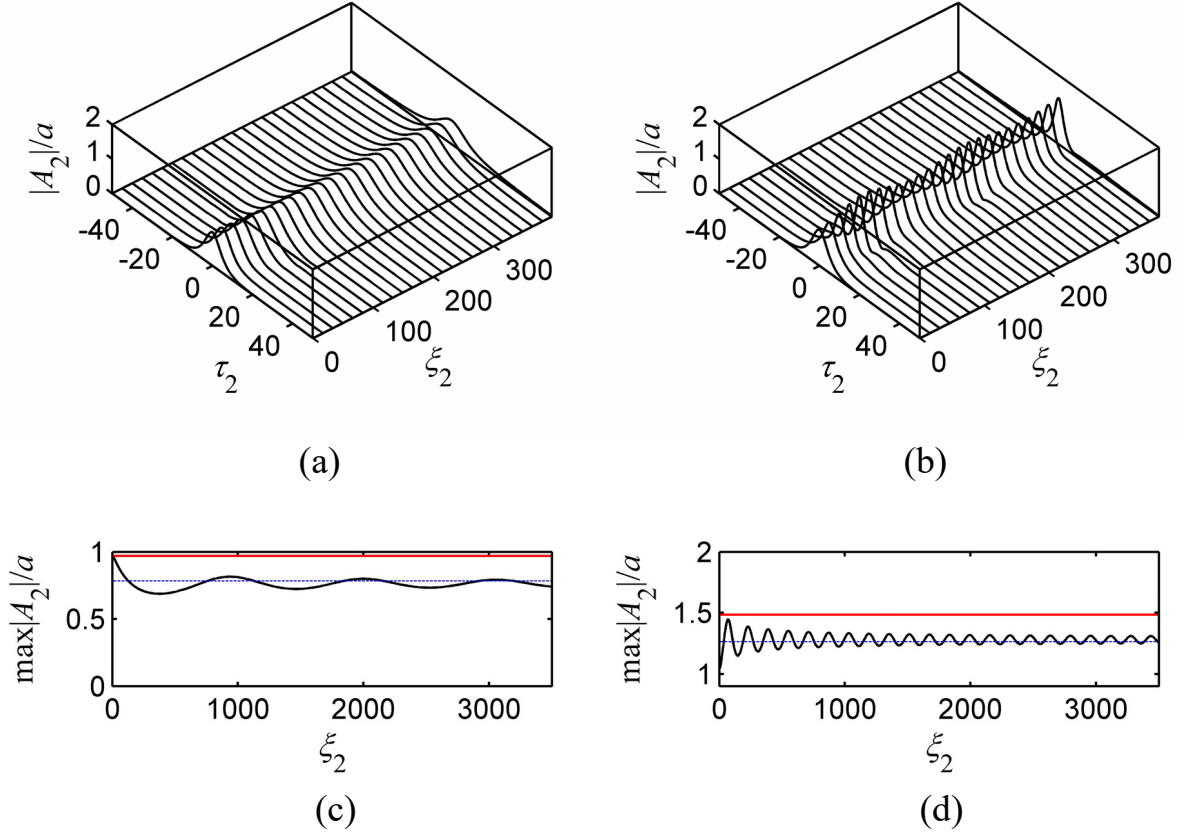

Fig. 7 (color online). Results of numerical simulations of the transmitted pulse evolution originated from the incident soliton with $k_1 a = 0.12$ and $\kappa_1 = 2$ within the NLS equation for $h_2/h_1 = 0.8$ (left) and $h_2/h_1 = 1.5$ (right). In frames (a) and (b) the evolutions of the envelope are shown, and in frames (c) and (d) the dependences of the envelope amplitudes on distance are shown. Horizontal dashed blue lines in frames (c) and (d) show the values of the predicted soliton amplitudes $a_1$; red solid lines show the value $(a_1 + a_{pw})/a$. Left: $a_1/a \approx 0.78$, $a_{pw}/a \approx 0.09$, right: $a_1/a \approx 1.26$, $a_{pw}/a \approx 0.22$.

Oscillations of amplitudes of pulse envelopes in the domain $\Omega_2$ become even more pronounced when two secondary solitons emerge; this is illustrated by Fig. 8. The two-soliton solution of the NLS equation, when solitons are located in one point, was considered by Satsuma & Yajima (1974) and Peregrine (1983). Such a solution is known as the bound solitons or bi-solitons. It represents a nonlinear beating between two solitons with the dimensional distance $L_b$, where

$$L_b = \frac{4\pi}{\alpha_2 \left(a_1^2 - a_2^2\right)}. \qquad (24)$$

In the process of beating, the amplitude of the envelope is limited by the sum of amplitudes of two solitons, $\max|A_2| \leq a_1 + a_2$. The two-soliton solution is one-humped at any time moment, if

$$\frac{a_2}{a_1} < \frac{3-\sqrt{5}}{2} \approx 0.38, \qquad (25)$$

and the minimum value of the envelope amplitude is bounded from below, $\max|A_2| \geq a_1 - a_2$ (we assume hereafter, that $a_1 > a_2$). According to Eq. (17), the condition (25) requires approximately $\mu < 2.12$; this is always fulfilled for the choice of parameters in the first domain $\Omega_1$ where $\kappa_1 = 2$, as illustrated by Fig. 8. Thus, the maximum value of two-soliton solution oscillates between the bounds:

$$a_1 - a_2 \leq \max|A_2| \leq a_1 + a_2. \qquad (26)$$

Blue dashed horizontal lines in Figs. 8c and 8d show the bounds of two-soliton solution given by Eq. (26). These bounds indeed capture rather well the limits of amplitude oscillations in the transmitted domain $\Omega_2$ far from the bottom step. The first local maxima in Fig. 8c,d occur a little bit later than it is predicted by the estimate (24).

The shapes of the numerically simulated envelopes and predicted by the exact two-soliton solution at the instants when the envelope amplitudes attain maximum and minimum values (shown by red dots in Fig. 8c and 8d) are compared in Figs. 8e and 8f. The envelope shapes obtained in simulations are given by thick lines, and exact solutions for the soliton amplitudes $a_1$ and $a_2$ calculated from Eq. (17) are shown by thin lines. As one can see, the envelope shapes agree quite well, but not perfectly. In the numerical simulations, the boundary condition at $\xi_2 = 0$, in contrast to the pure two-soliton solution, contains not only the envelope solitons but also a quasi-linear wavetrain. The wavetrain disperses and decays extremely slowly; its contribution to the solution can be seen in the right-side parts of Figs. 8e and 8f where the tails of exact and numerical solutions are shown in semilogarithmic axes.

The obtained solution in the domain $\Omega_2$ in the neighborhood of the point $x = 0$ can be interpreted as the nonlinear superposition of envelope solitons and a quasi-linear plane wave of effective amplitude $a_{pw}$. We assume that the dynamics of such a wavegroup possesses the characteristic features of a beating process well-known for linear waves with different frequencies. In the course of interaction, the components of the formation contribute to the wave maximum with their partial amplitudes *in a quasi-linear manner*. The linear superposition of partial amplitudes in nonlinear structures was observed in the collisions of many envelope solitons (Akhmediev & Mitzkevich, 1991; Sun, 2016), long-wave solitons (Slunyaev & Pelinovsky, 2016; Slunyaev, 2019), in the dynamics of breathers (Slunyaev et al., 2002; Slunyaev, 2006), and rational multi-breathers of the NLS equation (Wang et al., 2017). This underpins our assumption.

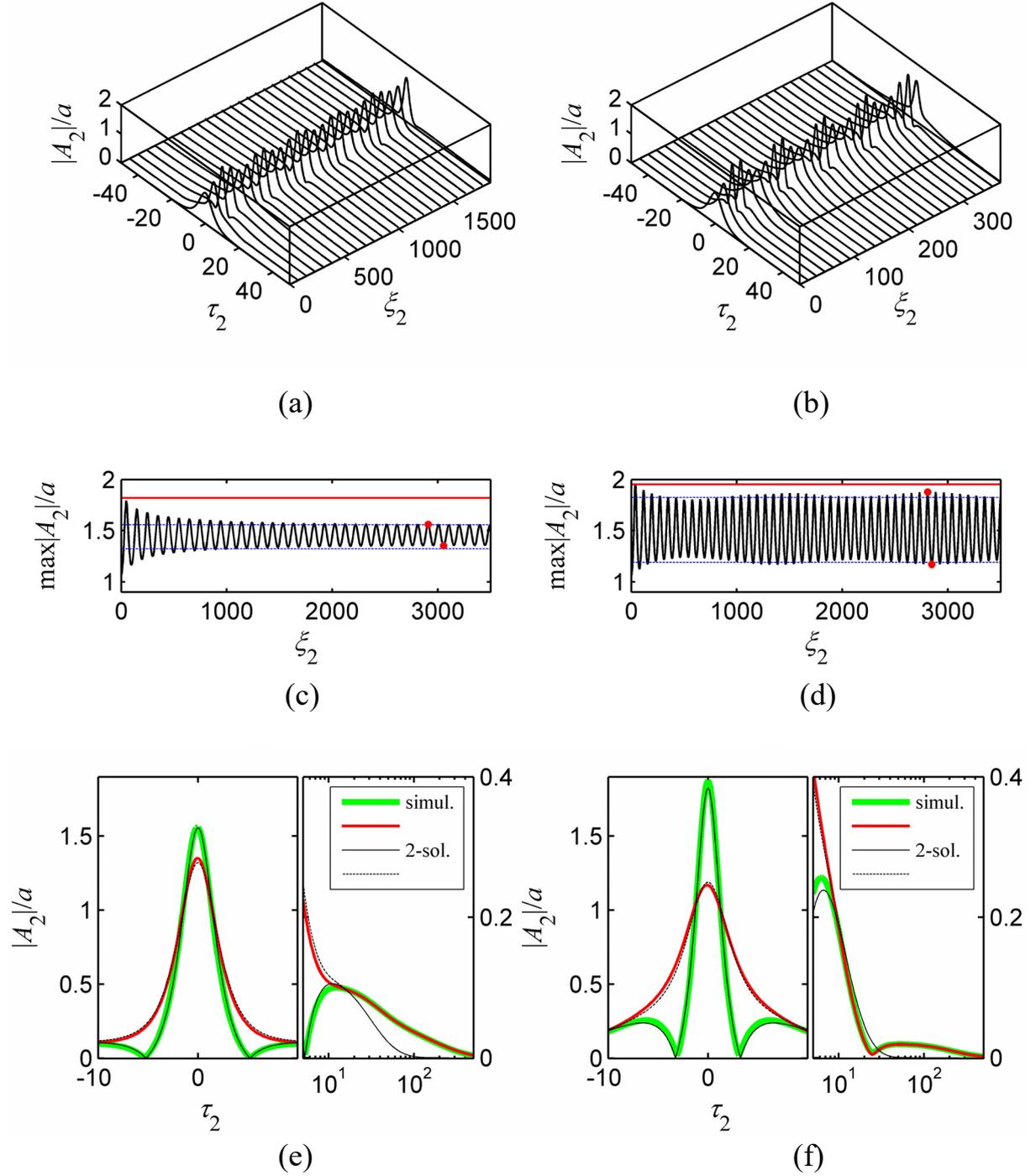

Fig. 8 (color online). Results of numerical simulations of transmitted pulse evolution originated from the incident soliton with $k_1 a = 0.12$ and $\kappa_1 = 2$ within the NLS equation for $h_2/h_1 = 2.5$ (left) and $h_2/h_1 = 10$ (right). In frames (a) and (b) the evolutions of the envelope are shown, and in frames (c) and (d) the dependences of the envelope amplitudes on distance are shown. In frames (c) and (d) the horizontal blue dashed lines show the values $(a_1 + a_2)/a$ and $(a_1 - a_2)/a$; red solid lines show the value $(a_1 + a_2 + a_{pw})/a$. In frames (e) and (f) the thick green and thick red lines represent the simulated envelope shapes at the distances where the wave amplitude attains local maximum and minimum; thin black and dotted lines correspond to the exact two-soliton solutions of the NLS equation (see the legend). Left: $a_1/a \approx 1.44$, $a_2/a \approx 0.12$, $a_{pw}/a \approx 0.26$, right: $a_1/a \approx 1.51$, $a_2/a \approx 0.32$, $a_{pw}/a \approx 0.13$.

In this study we assume that the wavefield amplitude at $x = 0_+$ can be presented as the linear superposition of a solitonic component and a background plane wave, $Ta = a_{sol} \pm a_{pw}$. The soliton component is the 'most asynchronous' superposition of secondary solitons with the amplitudes $a_1 > a_2 > \ldots > a_N > 0$. According to the results obtained in (Slunyaev, 2019), the amplitude of the solitonic component is $a_{sol} = a_1 - a_2 + a_3 - \ldots + (-1)^{N+1} a_N$. Then, the amplitude of the background wave $a_{pw} > 0$ can be estimated as:

$$a_{pw} = |a_{sol} - Ta|. \qquad (27)$$

The attainable maximum of the solution, $a_{max}$, is estimated as the amplitude of the 'most synchronized' wave,

$$a_{max} = a_1 + a_2 + \ldots + a_N + a_{pw}. \qquad (28)$$

The values of $a_{max}$ are shown in Figs. 7c,d and 8c,d by horizontal red solid lines. One can conclude that the values of $a_{max}$ provide remarkably accurate estimates of the maximum wave field in all cases. In the case when waves travel from deep to shallower water, solution (28) becomes trivial as it simply reduces to $a_{max} = Ta$.

Even more complicated and difficult for interpretation wavefield dynamics behind the step was observed in another series of simulations when the water depth in the domain $\Omega_1$ was shallower so that $\kappa_1 = 1.5$ (see Fig. 9). Firstly, condition (25) is failed when $h_2/h_1 > 2.47$; as a result, two-humped large wave groups can emerge (they can be seen in Figs. 9a,b); secondly, the lower limit in Eq. (26) becomes incorrect.

One more complication arises when $h_2/h_1 > 4.33$ due to the emergence of a third soliton (this case is shown in Fig. 9b,d). The horizontal red lines in Fig. 9c,d show the estimates for the maximal value of the wavefield as per Eq. (28); these estimates still provide rather good agreement with numerical data for the attainable wavefield maximum. Note that the wavefield absolute maximum is not the first in the series of maxima shown in Fig. 9c,d.

Formulae (27) and (28) can be re-written with the help of Eq. (17), as:

$$\frac{a_{pw}}{a} = T\left|\frac{N}{\mu} - 1\right|, \quad N = E\left(\mu + \frac{1}{2}\right), \qquad (29)$$

$$\frac{a_{sol}}{a} = \begin{cases} \dfrac{N}{S}, & \text{when } N \text{ is even;} \\ 2T - \dfrac{N}{S}, & \text{when } N \text{ is odd;} \end{cases} \qquad \frac{a_{max}}{a} = \begin{cases} T(2N-1) - \dfrac{N(N-1)}{S}, & \text{when } N > \mu; \\ T(2N+1) - \dfrac{N(N+1)}{S}, & \text{otherwise.} \end{cases}$$

As has been mentioned above, function $E(\cdot)$ returns the integer part of its argument; $N$ is greater than $\mu$ when the fractional part of $\mu$ is greater than 0.5.

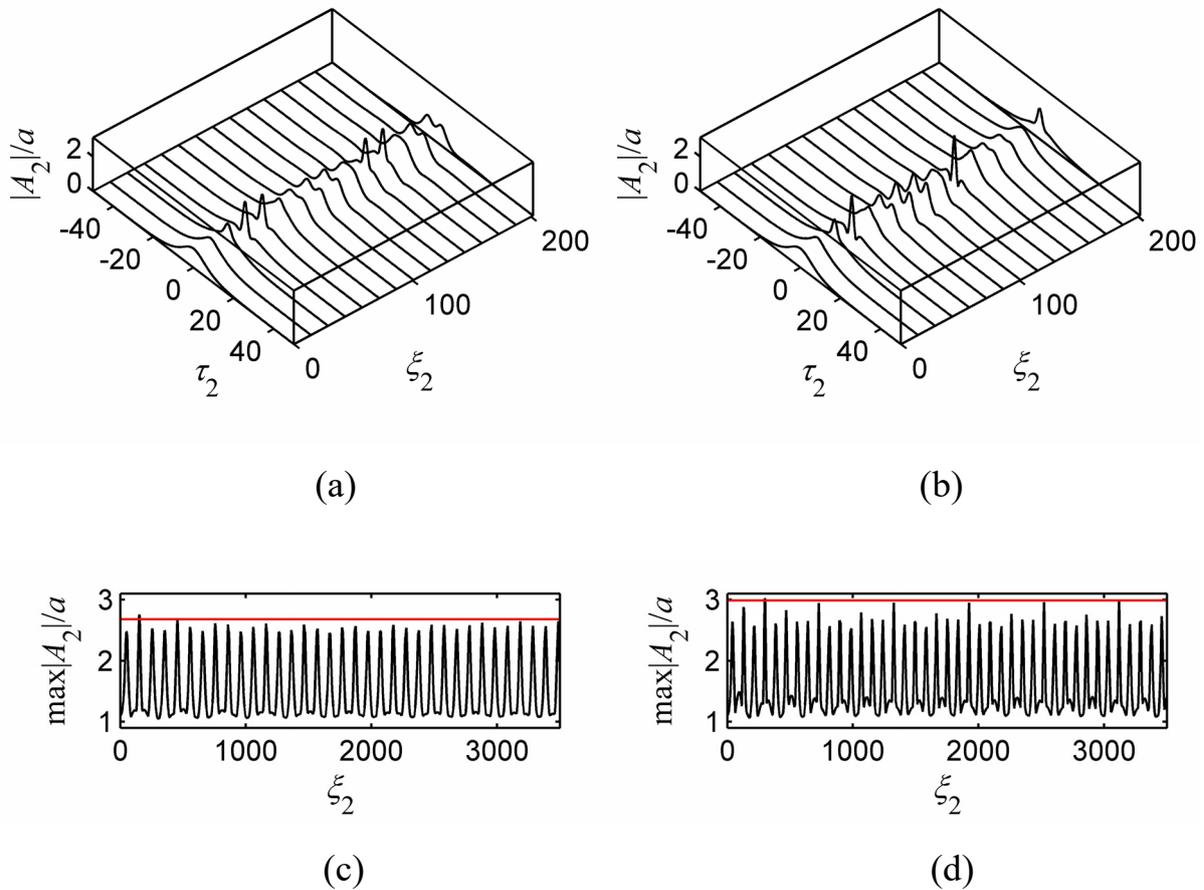

Fig. 9 (color online). The same as in Fig. 7, but for the transition from the domain $\Omega_1$ where $\kappa_1 = 1.5$ to the domain $\Omega_2$ where $h_2/h_1 = 3.33$ (left) and to $h_2/h_1 = 13.3$ (right). The horizontal red line shows the value of $a_{max}/a$ according to Eq. (28). Left column: $a_1/a \approx 1.71$, $a_2/a \approx 0.80$, $a_{pw}/a \approx 0.17$, right column: $a_1/a \approx 1.76$, $a_2/a \approx 0.95$, $a_3/a \approx 0.15$, $a_{pw}/a \approx 0.13$.

The results of a series of numerical simulations for the depths $\kappa_1 = 2$ and $\kappa_1 = 1.5$ are compared in Figs. 10a and 10b, respectively, with the estimate (29). We also reproduce the curves for the soliton amplitudes $a_n$ from Fig. 6 and plot the values of $a_{pw}$. Formula (29) describing the wave extremes $a_{max}$ in the domain $\Omega_2$ agrees well with the numerical simulations (the actual extremes, $\max|A_2|$, are plotted by circles) despite a complicated broken dependence for $a_{max}$. According to Fig. 10, amplitudes of nonlinear wavetrains travelling from the domain $\Omega_1$, where $\kappa_1$ varies between 1.5 and 2, towards a deeper domain $\Omega_2$, can increase in 2–3 times.

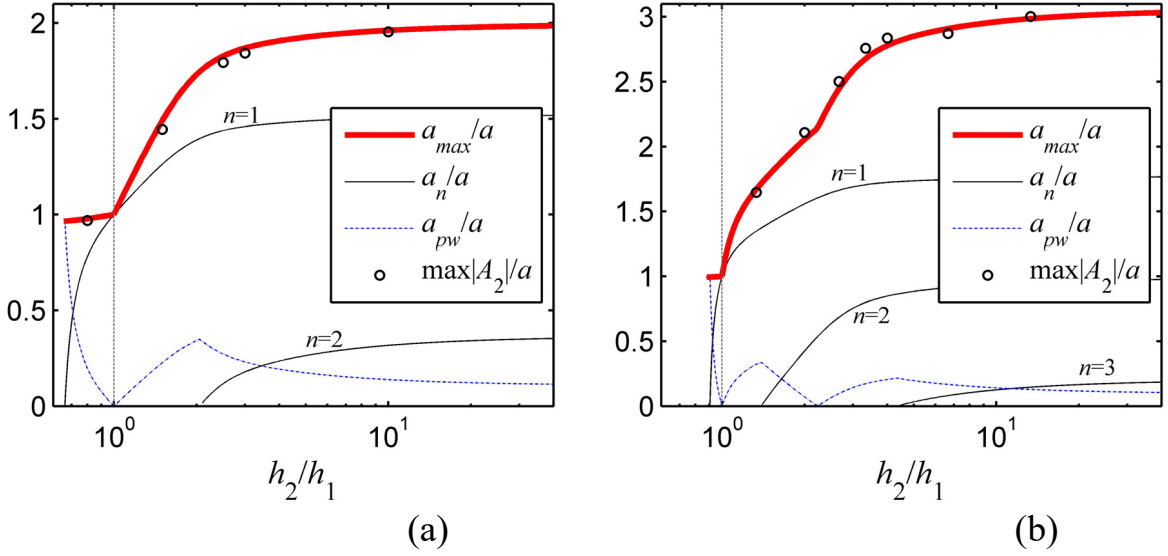

Fig. 10 (color online). Amplification factors of envelope solitons for the cases $\kappa_1 = 2$ (a) and $\kappa_1 = 1.5$ (b). The data of numerical simulation of the maximum value of $|A_2|/a$ obtained within the NLS equation are compared with the estimated maximum wave amplitude in the domain $\Omega_2$ $a_{max}/a$. The normalized soliton amplitudes $a_n/a$ and the estimated value of the background linear wave $a_{pw}/a$ are also shown.

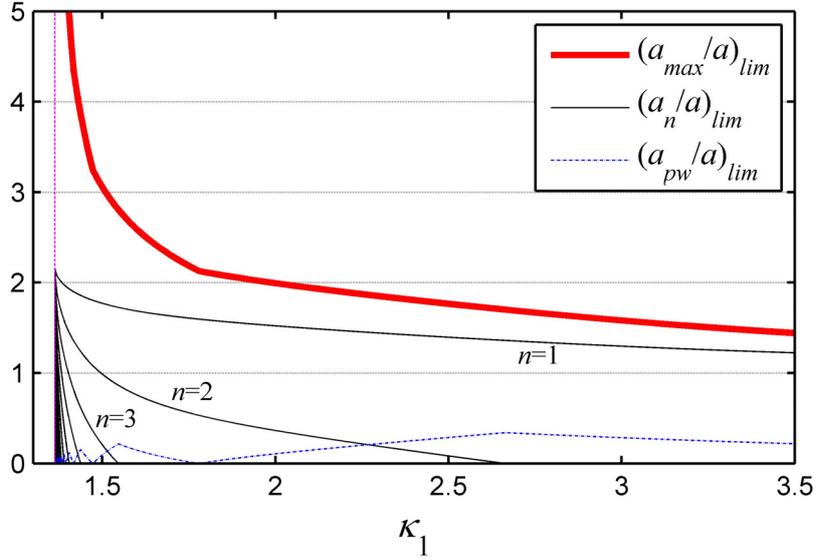

Fig. 11 (color online). Amplification factors of envelope solitons in the limiting case when the water depth in the domain $\Omega_2$ is infinite.

The process of beating between the emerged secondary solitons and the background quasi-linear wave results in the remarkably greater enhancement than the one predicted by the linear theory (see Fig. 4) or based on the consideration of non-interacting solitonic component only (see Fig. 6). The maximum amplification is achieved when the water in the domain $\Omega_2$ is infinitely deep. Solution for this limiting case, $\kappa_2 \to \infty$, depends on the scaled water depth in the domain $\Omega_1$ and is shown in Fig. 11. In the figure the curves of the wave maximum, of the soliton amplitudes and of the quasi-linear wave component are plotted, similar to Fig. 10. While the amplitude of the leading soliton $a_1$ remains finite when $\kappa_1 \to 1.363$, the maximum attainable amplitude $a_{max}$ grows infinitely due to the infinite number of secondary solitons which can superimpose ($\mu$ tends to infinity as $\alpha_1$ approaches zero, this asymptote is shown by the vertical dashed line in Fig. 11).

Finally, we reiterate that the wave amplification $a_{max}/a$ obtained within the NLS theory does not depend on the amplitude of the incident soliton.

## 5. Numerical simulation of soliton transformation within the primitive set of hydrodynamic equations

The problem of transformation of an envelope soliton at a bottom step has been studied also within the primitive set of equations for the potential flow of ideal fluid in the planar geometry (see Fig. 1). A numerical model based on the High Order Spectral Method (HOSM) (Dommermuth & Yue, 1987; West et al, 1987) and applicable to a smooth variable topography was used (Gouin et al, 2016). The code takes as a basis the open-source solver HOS-ocean (Ducrozet et al, 2016). After a careful analysis of code convergence, the order of nonlinearity in the computations was fixed to 8 (i.e., up to nine-wave nonlinear interactions were considered). A lengthy spatial domain of the total size of about 900 wavelengths was simulated imposing periodic boundary conditions with the absorbing zones close to the boundaries that effectively lead to the non-reflecting boundary conditions. When it was necessary, the size of the domain $\Omega_2$ was further extended. The capability of the code is limited by moderate and smooth variations of the water depth. The bottom step in the middle of the computation domain was specified by a smooth transitional profile using a piece of a sinusoid of the horizontal size $\Delta$, which was typically as long as three dominant wave lengths, $k_1\Delta = 6\pi$. The validity of the numerical model has been studied thoroughly by Gouin et al. (2016). Different test-cases can be found with results validated with the bottom slopes up to 0.5 and water depth variation up to 75%. The study has demonstrated that the convergence rate of the numerical scheme reduces when the bottom variation and bottom slope increases. However, the method allows an accurate solution up to the aforementioned limits. In the numerical results presented in this section, the bottom slope varies in the range from 0.033 to 0.38, and the relative water depth variation in the range from 11% to 60%. The numerical parameters of the study have been chosen after a careful convergence analysis which secures accurate computations. In particular, the energy integral was conserved with an accuracy of at least 0.1%.

The HOSM code calculates the wave evolution in time, in contrast to the previously used spatial NLS equations (2). The initial condition was specified in the form of the exact envelope soliton solution of the NLS equation (2) for $j = 1$ located far enough in front of the bottom step. The scaled amplitudes, which have the meaning of the wave steepness, were $k_1 a = 0.06$ and $k_1 a = 0.12$ in different runs. Since the initial conditions were not exact solutions of the hydrodynamic equations, their evolution at the early stage was accompanied by a radiated trailing wave that ran behind a soliton prior it approached the bottom step. As a result, the solitary nonlinear wavetrain had a well-formed shape when it entered the zone where the depth becomes changing. Since the actual incident wavetrain was slightly different from the initial pulse, the amplitude of the incident soliton was estimated as a half of the wave heights averaged over 20 wave periods prior the depth change, $a = <A_{cr} + A_{tr}>/2$, where $A_{cr} = \max(\eta_1)$ and $A_{tr} = -\min(\eta_1)$ are the momentary maximum and minimum of the wavetrain, and the angular brackets denote an average value. These slightly reduced values of $a$ (see the example below in Fig. 15) were used for the comparative simulations of the NLS equation.

The dynamics of the transition of a solitary group with the steepness $k_1 a = 0.12$ from the domain $\Omega_1$ where $\kappa_1 = 2$ to the domain $\Omega_2$ with the depth jump $h_2/h_1 = 2.5$ is illustrated in Fig. 12, the conditions are similar to the ones shown in the left column of Fig. 8. The evolution in time, simulated by the HOS code, was recalculated to the evolution in space, using the linear wave group velocity in the domain $\Omega_2$, $\xi_2 = k_2 C_2 t /(2\pi)$. The solid curves represent the normalized envelope amplitudes, $\max|A_2|$, of the wavefield calculated from the simulation within the NLS equation (line 1) and from the Euler equations (line 2). The latter is estimated through the momentary wavetrain maximum and minimum, $(A_{cr} + A_{tr})/2$, while the interval between the values $A_{cr}$ and $A_{tr}$ is filled with color. The wavetrain maxima and minima are not equal in absolute value due to the vertical wave asymmetry caused by nonlinearity. They experience fast oscillations, owing to the difference between the velocities of the individual wave and wavetrain. The horizontal lines, like in Fig. 8c, show the normalized values of $a_{max}$, $a_1 + a_2$ and $a_1 - a_2$.

In the simulation of the Euler equations the envelope amplitude, which is amplified due to the depth increase, is a little bit smaller compared to the simulation with the NLS equation. The rate of the wave enhancement right after the bottom step is somewhat slower, and the period of oscillations is longer in the simulations within the HOSM. The dynamics of self-modulation of a quasi-periodic wave group was observed in the numerical simulations within the Euler equations; the dynamics looks qualitatively similar to the results of simulation within the NLS equations, the bands of oscillations are similar as well. Due to the vertical wave asymmetry, the amplification of wave crests can exceed the value predicted by the approximate theory as per Eq. (29), though the maximum wave half-height $(A_{cr} + A_{tr})/2$ is smaller than the envelope amplitude within the weakly nonlinear framework. Due to the non-integrability of the Euler equations, one can expect that the soliton-like groups can be changed in the course of inelastic interactions. Therefore, in the longer-term evolution the solution can significantly differ from that obtained within the NLS equation.

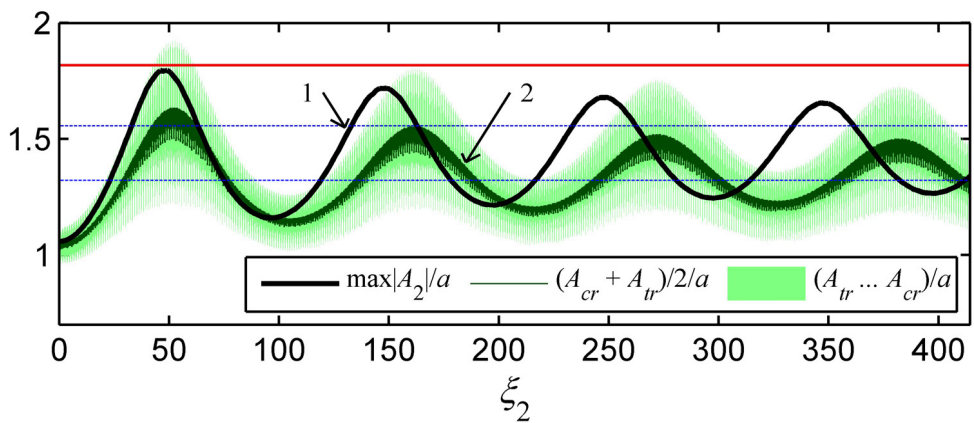

Fig. 12 (color online). The dependence of wave amplitude on distance obtained in the direct numerical simulation compared with the result of simulation of the NLS equation. The simulation conditions are similar to those shown in Fig. 8a,c.

The upper parts of Figs. 10a,b are reproduced in Figs. 13a,b correspondingly, where maximum wave enhancements observed in the series of numerical simulations within the Euler equations are also shown. The cases when the solitary groups originally propagate over the depths with $\kappa_1 = 2$ and $\kappa_1 = 1.5$ are presented separately in frames (a) and (b), respectively. The vertical bars show the bands of maximum amplification of the wavetrain maxima and minima. They represent the intervals [max($A_{tr}$)/$a$, max($A_{cr}$)/$a$], where the maxima were calculated in the domain $\Omega_2$. The cases of two different wave steepness of the incident soliton, $k_1 a \approx 0.06$ and $k_1 a \approx 0.12$, have been studied (see the legends in Fig. 13). The results of the weakly nonlinear theory, the estimate $a_{max}/a$ as per Eq. (29) (the thick solid curves) and the amplifications obtained in the direct numerical simulation of the NLS equation (circles), are reproduced for the comparison. The largest soliton amplitude $a_1/a$ (the thin solid curve) is also shown for the reference.

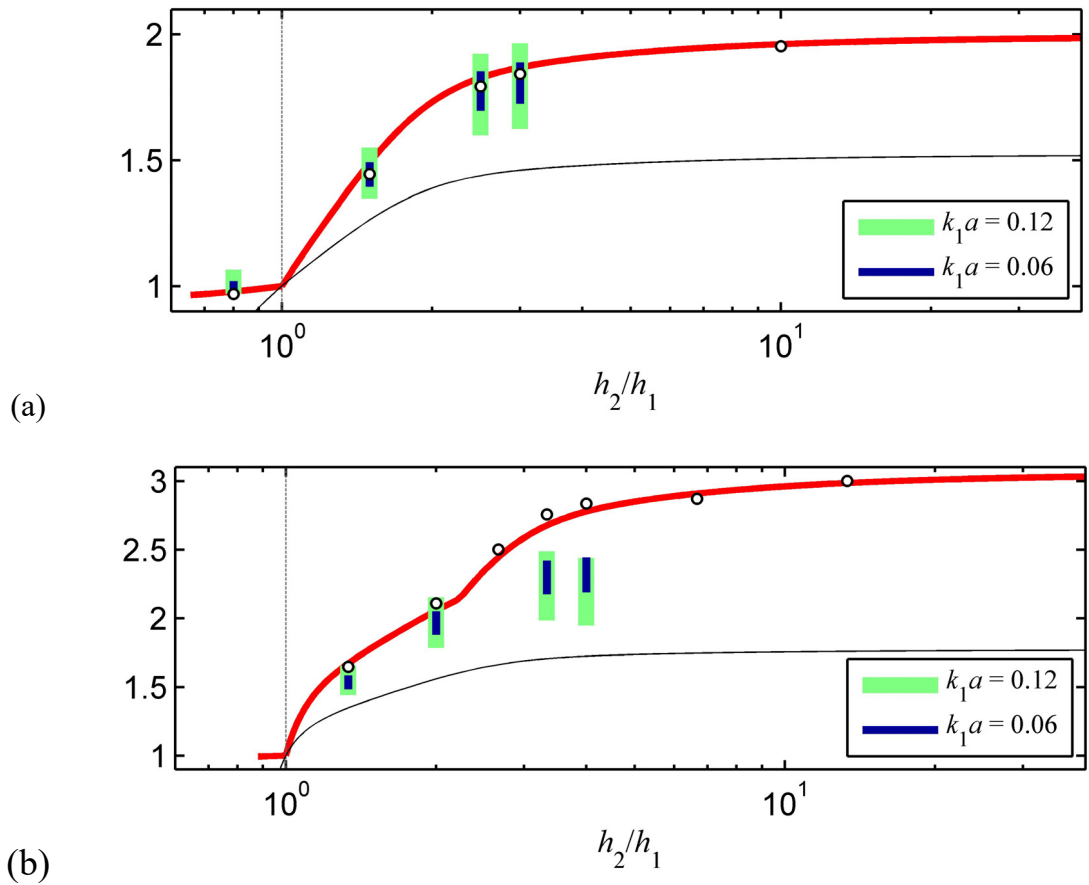

(a)

(b)

Fig. 13 (color online). Maximum wave amplification for the envelope soliton transition from the domain $\Omega_1$ to $\Omega_2$ with the depth parameters $\kappa_1 = 2$ (a) and $\kappa_1 = 1.5$ (b). The bars represent the result obtained within the numerical simulation of the HOSM for $h_2/h_1 = 0.8, 1.5, 2.5, 3$ (a); and $h_2/h_1 = 1.3, 2, 3.3, 4$ (b). The thick solid line shows the solution provided by the weakly nonlinear theory $a_{max}/a$ as per Eq. (29), and the open circles represent the result of numerical simulation within the NLS equation.

As one can see from Fig. 13a for $\kappa_1 = 2$, the maximum wave amplification in the simulations of the HOSM follows the theoretical curve. The bars which correspond to the simulations of small-amplitude waves agree better with the weakly nonlinear solution, whereas steeper waves result in longer bars (because the wavetrains possess stronger vertical asymmetry), which are still described reasonably well by the analytic curve for $a_{max}$. The agreement is noticeably worse when the domain $\Omega_1$ is shallower so that $\kappa_1 = 1.5$ (see Fig. 13b), though the amplification is greater than in the previous case. In the case with $\kappa_1 = 2$, the NLS theory predicts the emergence of a second soliton after the depth transition only when $h_2/h_1 = 3$; then $a_1/a \approx 1.46$ and $a_2/a \approx 0.18$, and the second soliton is small. In another case with $\kappa_1 = 1.5$, two solitons emerge in all simulations shown in Fig. 13b for $h_2 > h_1$; in the case $h_2/h_1 = 4$ the soliton amplitudes are evaluated according to Eq. (17) are as much as $a_1/a \approx 1.72$ and $a_2/a \approx 0.85$. It is important to note that the employed method of simulation of the Euler equations becomes less accurate when the depth drop $h_2/h_1$ becomes too large. Therefore, the deviation of two rightmost bars in Fig. 13b from the theoretical thick line can be caused by the shortcomings of the numerical code.

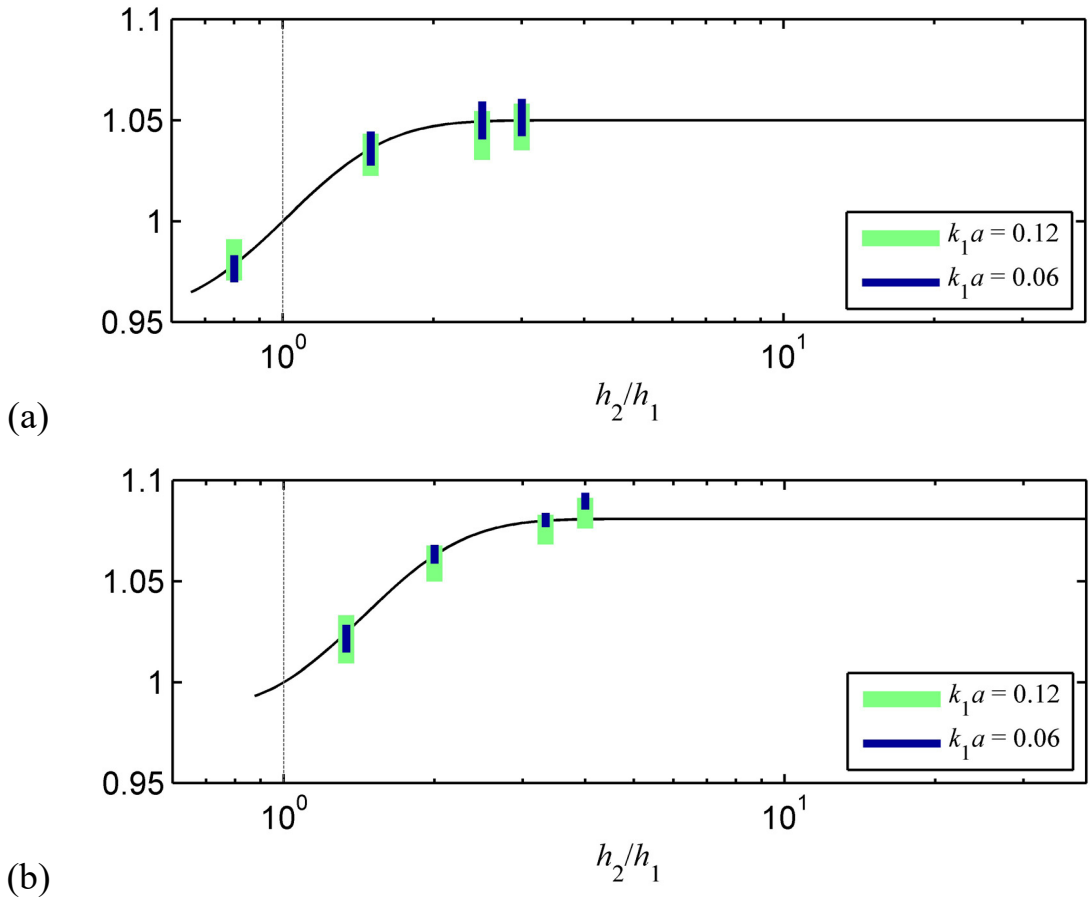

(a)

(b)

Fig. 14 (color online). Estimation of the transmission coefficient $T$ in the direct numerical simulations within the Euler equations shown in Fig. 13a,b (bars). Solid lines show solution (19) obtained in the linear approximation.

In our theoretical consideration presented in Sec. 3, it was assumed that the wave transformation over the bottom step is much quicker in comparison with the characteristic times of manifestation of nonlinear and dispersion effects; this legitimates the use of the transmission coefficient $T$ (19) found within the linear theory for monochromatic waves. This coefficient was verified in (Giniyatullin et al., 2014; Kurkin et al., 2015a) by direct numerical simulation of the primitive set of hydrodynamic equations for small amplitude waves. In Fig. 14 we compare theoretical formula (19) with the estimation of the quick wave amplification in the direct numerical simulations within the Euler equations. To this end, we consider the evolution of the maximum wave amplitude $(A_{cr} + A_{tr})/2$ as the function of time. This quantity experiences a quick change when the group passes over the bottom step, as shown in Fig. 15. The time moments when the quick change of wave amplitude occurs was selected by eye and the wave amplitudes were calculated then by averaging over 20 wave periods in front of the step ($a_i \pm \Delta a_i$) and 5 wave periods past the step ($a_t \pm \Delta a_t$) (the standard deviations were used as the confidence intervals). Then, the numerically estimated transmission coefficients $T$ were plotted in Fig. 14 as the bands $[(a_t - \Delta a_t)/(a_i + \Delta a_i), (a_t + \Delta a_t)/(a_i - \Delta a_i)]$ for two different amplitudes of an incident soliton. Solution for the linear uniform waves as per Eq. (19) is shown in Fig. 14 by solid lines. As one can see, there is a good agreement between the coefficient $T$ in Eq. (19) and the results of direct numerical simulation for strongly nonlinear modulated waves. Some discrepancy between these data is observed for $\kappa_1 = 1.5$ and $h_2/h_1 = 4$ but this can be caused by the numerical artifacts as has been mentioned above.

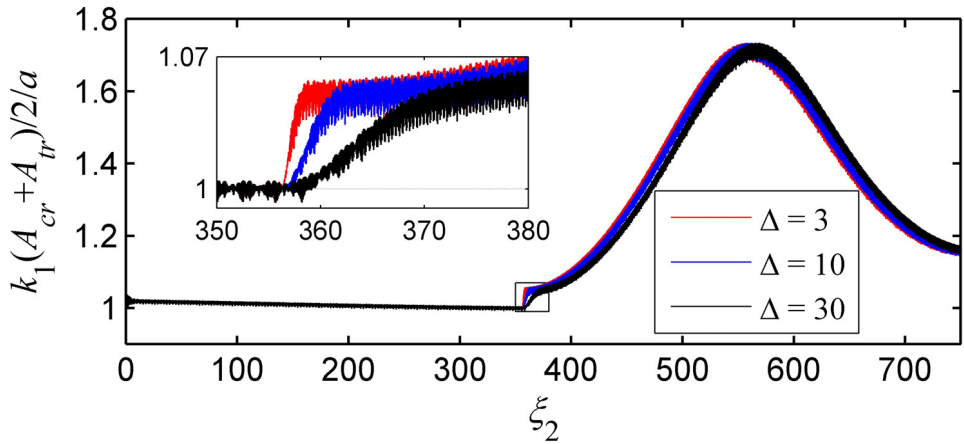

Fig. 15 (color online). Dependence of the maximum wave amplitude as a function of distance in the direct numerical simulation within the primitive set of hydrodynamic equations for $\kappa_1 = 2$, $h_2/h_1 = 2.5$ and the incident envelope soliton steepness $k_1 a \approx 0.06$. Three curves show the numerical results for different lengths of the bottom step $\Delta = 3$, 10, and 30 wavelengths. The process of pulse passing over the bottom step is shown in the magnified insertion.

In contrast to the previously used NLS theory, the numerical simulations within the primitive set of hydrodynamic equations were performed for the bottom profile, when the depth change was prescribed by a smooth function with the horizontal size $\Delta$ of three dominant wavelengths. In Fig. 15 we present a comparison of the evolution of solitary wavetrain for three values of Δ consisting of 3, 10, and 30 wavelengths for the following parameters: $\kappa_1 = 2$, $h_2/h_1 = 2.5$, $k_1 a \approx 0.06$. As one can see, the characteristic length Δ of the bottom step has practically no influence on either the quick amplitude increase in the very vicinity of the bottom step (see the inset), or the pulse evolution past the step.

## 6. Conclusion

In this work the transformation of envelope solitons on a bottom step was studied analytically within the weakly nonlinear theory for slowly modulated waves using the transformation coefficient for linear waves suggested in (Giniyatullin et al., 2014; Kurkin et al., 2015a). This was also studied numerically using the simulations within the nonlinear Schrödinger equation and primitive set of hydrodynamic equations for the potential flow. The wide range of water depths was considered, from the relatively shallow water up to infinitely deep water. The wave parameters were chosen such that the effects of nonlinearity and dispersion could be balanced to support long-living solitary wave groups. Such structurally stable nonlinear groups with high amplitudes and steepnesses up to $k_0 A_{cr} \approx 0.3$ were observed in laboratory simulations (Slunyaev et al., 2013; 2017). It was confirmed that soliton-like patterns embedded into irregular wave trains can propagate for a long distance (Viotti et al., 2013; Wang et al., 2020; Slunyaev, 2021), therefore they represent both the theoretical and practical interest.

When an envelope soliton propagates from the deep to shallower water, its amplitude decreases right after the step and later on. The transmitted wavetrain can completely disperse if the depth in the shallower domain is too small. When an envelope soliton enters a deeper domain, its amplitude right after the depth change increases, and the transmitted wavetrain can split into one or several envelope solitons and a trailing quasi-linear wave. The wave pattern forms a bound state and interacts in a linear fashion; in the result, it can produce a high-amplitude wave in the course of propagation.

The observed wave amplification occurs after a quick transformation of a quasi-linear wave train at the bottom step and consequent nonlinear evolution of the transmitted wave packet. Its evolution resembles the effect of modulational instability triggered by head-one currents or strong winds (Onorato et al., 2011; Slunyaev et al., 2015). As the result, the leading envelope soliton which is generated after the depth drop theoretically can have the amplitude of more than two times greater than the amplitude of incident wave group. Within the integrable NLS equation, the solitons represent the asymptotic solution in the transmitted wavefield after the bottom step. As for the third effect, a large number of secondary solitons can emerge from the transmitted wave group if the depth drop is sufficient. They can have significant amplitudes and create big-amplitude waves in the course of propagation behind the bottom step due to the constructive interference between them and with the quasilinear residual.

The analytic formula for the maximum wave amplification obtained within the framework of the nonlinear Schrödinger equation is proposed in this paper; it can provide a very big value at a relevant relationship between the depth in two domains. The numerical simulation within the NLS equation confirms correctness of the theoretical description. The results of the direct numerical simulation within the primitive set of hydrodynamic equations agree well with the weakly nonlinear theory, though demonstrate noticeable deviation when the depth change is too big; this can be explained by limitations of the adapted numerical scheme based on the High-Order Spectral Method. The maximum wave amplification attained in the series of numerical simulations exceeded factor two, therefore, the wave trains in such case can be referred to rogue waves (Kharif et al., 2009).

To conclude, when an envelope soliton enters a deeper region, the amplification of the soliton amplitude is stronger than the one of linear waves; the transmitted wave group can disintegrate into several solitons and a quasi-linear trailing wave. The interaction between the solitons and the quasi-linear wave further enhances the amplification considerably. Note that though we limited the consideration by the conditions when envelope solitons may exist in both water domains requiring that $k_j h_j > 1.363$ for $j = 1, 2$, the restriction can be slightly weakened for the first domain from which the envelope soliton approaches the bottom step. The nonlinear mechanisms of wave amplification add to the linear wave enhancement if an envelope soliton emerges right after the step. Similarly, the incident wavetrain can have the shape different from the NLS envelope soliton. It should be also mentioned that the threshold of the modulational instability $kh = 1.363$ alters when higher order effects are considered. This problem was analyzed in (Slunyaev, 2005; Agafontsev, 2008), where the shift of the instability boundary to shallower water was concluded (in contrast to some other works, see a discussion in (Slunyaev, 2005)). It was shown in (Grimshaw & Annenkov, 2011), that solitary wave packets which propagate shoreward can either penetrate into shallow water, $kh < 1.363$, or even not reach this depth depending on the group characteristics.

The developed theory based on the 1D NLS equation can describe the 3D problem when the wavevectors of the carrier and envelope waves are at an angle to each other within the band of instability (slanted envelope solitons, see (Chabchoub et al., 2019). The effective 1D NLS equation for the envelope will have the form of Eq. (2) with the modified coefficients $C_{1,2}$ and $\beta_{1,2}$ and with the similar statement of the problem. Envelope solitons are known to be unstable with respect to transverse perturbations (Zakharov & Rubenchik, 1974; Ablowitz & Segur, 1979) therefore, they cannot propagate for a long distance in the open sea. However, narrow channels or waveguides formed by lateral inhomogeneities can stabilize the soliton (see (Shrira & Slunyaev, 2014)) and hence, increase feasibility of the described effects in nature.

Though a hydrodynamic example is considered in this work, the obtained qualitative understanding obviously has a broader application to other nonlinear media which support the propagation of envelope solitons. The analytic description (29) should remain applicable if the nonlinear Schrödinger equation serves as the leading-order approximation. It can be a challenge to experimentalists to validate the results obtained in this paper by the data of laboratory modelling or field observations. Such works were conducted for the observation of transformation of KdV solitons on bottom steps (Seabra-Santos et al., 1987; Losada et al., 1989).

**Acknowledgements**

G.D. acknowledges the support by the International Emerging Actions program of CNRS (grant No. IEA00464). A.S. was partially supported by Laboratory of Dynamical Systems and Applications NRU HSE, of the Ministry of Science and Higher Education of the Russian Federation Grant 075-15-2019-1931; and by the Russian Foundation for Basic Research (grant No. 21-55-15008). A.S. also acknowledges the support from the International Visitor Program of The University of Sydney in 2019 and is grateful to the School of Sciences, University of Southern Queensland for warm hospitality. Y.S. acknowledges the funding of this study provided by the State task program in the sphere of scientific activity of the Ministry of Science and Higher Education of the Russian Federation (project No. FSWE-2020-0007) and the grant of President of the Russian Federation for the state support of Leading Scientific Schools of the Russian Federation (grant No. NSH-2485.2020.5). The paper was completed when Y.S. was working under the SMRI visiting program at The University of Sydney in March–April 2021. Y.S. is thankful for the provided grant and hospitality of SMRI staff.

## Data Availability

The data that support the findings of this study are available from the authors upon reasonable request.